\documentclass[unnumsec,webpdf,contemporary,large]{oup-authoring-template}%

\onecolumn 

\graphicspath{{Fig/}}

\theoremstyle{thmstyleone}%
\theoremstyle{thmstyletwo}%
\theoremstyle{thmstylethree}%

\usepackage{longtable}
\usepackage{amsmath}
\usepackage{amsfonts}
\usepackage{graphicx}
\usepackage{float} 
\usepackage{xcolor}

\begin{document}

\journaltitle{Journal Title Here}
\DOI{DOI HERE}
\copyrightyear{2022}
\pubyear{2019}
\access{Advance Access Publication Date: Day Month Year}
\appnotes{Paper}

\firstpage{1}


\title[Crime in Proportions]{Crime in Proportions: Applying Compositional Data Analysis to European Crime Trends for 2022}
\author[1]{Onur Bat\i n DO\u{G}AN\ORCID{0000-0003-4632-2884}}
\author[1,$\ast$]{Fatma Sevin\c{c} KURNAZ\ORCID{0000-0002-5958-7366}}


\address[1]{\orgdiv{Department of Statistics}, \orgname{Yildiz Technical University}, \orgaddress{\street{Davutpasa Campuss}, \postcode{34220}, \state{Istanbul}, \country{Turkey}}}

\corresp[$\ast$]{Corresponding author. \href{email:email-id.com}{fskurnaz@yildiz.edu.tr}}


\received{Date}{0}{Year}
\revised{Date}{0}{Year}
\accepted{Date}{0}{Year}



\abstract{This article investigates crime patterns across European countries in 2022 using Compositional Data Analysis (CoDA) to address limitations of traditional statistical approaches in dealing with the relative nature of crime data. Recognizing crime types as components of a whole, we employ CoDA to explore relationships between different crime categories while respecting their inherent interdependencies. The study utilizes k-means clustering to group countries based on their crime profiles, identifying three distinct clusters largely aligning with geographical locations. This clustering is visualized through t-SNE and geographic mapping, revealing regional similarities. Further analysis using Robust Principal Component Analysis on identified crime clusters reveals insightful relationships between specific crime types, such as homicide, smuggling, and financial crimes, and how their prevalence varies across countries. The findings reveals distinct crime patterns across Europe, highlighting regional commonalities while also highlighting divergences like Norway and Latvia that deviate from their expected geographical classifications. Moreover, the study  identifies specific crime groups; for example, it pairs countries high in corruption and smuggling, such as Austria, with those countries that exhibit a higher relevance to homicide and smuggling, such as Luxembourg. It also points to the presence of financial crimes like fraud in countries such as Romania and Estonia.}


\maketitle

\section{Introduction}
Crime and its attendant risks are at least one of the greatest challenges human societies have today. Some of the most important phenomena in European society are related to crime and consequently on public safety, economic stability, or social cohesion; thus this field has attracted the attention of policymakers and researchers. Addressing feelings of unsafety in an area might contribute to better general health of residents \cite{ruijsbroek2015social}. Crime, in addition to direct costs like property losses to damages to loss impacts indirectly through lower quality of life, heightened fear of victimization and disruptions in the long-term community well being \cite{Dijk_2008} and; furthermore, is an expensive business. Consequently, the economic toll of crime must be considerable and estimates indicate that the gambling costs in the EU approach billions of euros per annum \cite{brand2000}. There is a need for empirical knowledge on the distribution and composition of different criminal entities across the European countries in order to make realistic policy and intervention decisions.

Typically, statistical analyses of crime data have been conducted using traditional methods, for example, linear regression and correlation studies. Linear regression models have been used as an estimator of the association between crime rates and socioeconomic factors, such as unemployment or income inequality \cite{Fajnzylber2002}. In a similar fashion, multivariate techniques like cluster analysis and factor analysis have been applied to characterize or characterize groups of crime types \cite{xu2005crimenet}. However, while these are certainly effective traditional methods they often based on absolute crime counts or rates; which can present primarily with several methodological challenges \cite{filzmoser2010bivariate}. Crime data being compositional is inherent in that it is for sure some sum and parts-of-them — for example, the total number of all types of crime in a region/country (e.g., violent, corruption, burglary) overall is a sum (often expressed in proportions), and relative information between categories \cite{egozcue2009reply}. It is only feasible to analyse compositions directly in their geometry, but it is resource-intensive and may return uncertain results. Given that standard statistical methods are designed to be applied in Euclidean space, it would be more appropriate to create transformations that convert compositional data into real space for a meaningful multivariate analysis \cite{filzmoser2018applied}. Compositional data analysis (CoDa) provides a consistent framework to overcome the limitations of traditional approaches when analyzing data that is parts of a whole. By transforming into appropriate coordinates, CoDA removes the singular covariance matrix caused by the constant-sum constraint and prevents the use of methods that depend on regular covariance matrices \cite{nesrstova2023simple}. This ensures that statistical analyses are valid and interpretable. This means that an increase in one type of crime must necessarily result in a relative decrease in the proportions of other crime types, regardless of their absolute counts. Such interdependencies between components violate the assumptions of traditional statistical methods, which typically treat variables as independent. By treating crime data as compositional, we can focus on the relative relationships between crime types, enabling meaningful comparisons and interpretations \cite{pawlowsky2015modeling}. 

The rest of the article is organized as follows: First, we explain our methodology and give main ideas behind the techniques shortly. Specifically, we analyze the observations, or countries, using clustering techniques that allow us group countries with similar crime profiles together, followed by t-SNE visualization in order to make exploration and interpretation of the spatial distribution of these clusters possible. The Q-mode clustering in the second part, while focusing on the variables-that is, the crime types-discovers groups of related crimes to which PCA is applied within each cluster to outline the pattern and relationships hidden beneath the surface. Then, these will be discussed in terms of their implications for crime prevention, policymaking, and comparisons across countries, summarizing insights and suggesting further areas of research.

\section{Materials and methods}
Data preparation and the entire statistical analysis were carried out using R Statistical Software, version 4.3.1 \cite{R23}. The \textit{robCompositions} package \cite{templ2011robcompositions} was used for compositional data analysis, including imputation and PCA, while the \textit{Rtsne} \cite{rtsne2015package} was utilized to perform t-SNE visualizations. The \textit{cluster} package \cite{cluster2022package} provided clustering algorithms, also, the \textit{mclust} package \cite{mclust2023package} was employed for Gaussian Mixture Models (GMM). For graphical representation, the \textit{dendextend} \cite{dendextend2015package} and \textit{ggplot2} \cite{ggplot22016package} packages were used. Additionally, the packages \textit{countrycode} \cite{countrycode2018package}, \textit{rnaturalearth} \cite{rnaturalearth2023package}, \textit{rnaturalearthdata} \cite{rnaturalearthdata2024package}, \textit{sf} \cite{sf2023package}, and \textit{ggrepel} \cite{ggrepel2024package} were applied to create detailed mapping visualizations.

\subsection{Data Collection and Preprocessing}

Dataset of ``Corruption \& Economic Crime" but the categories ``Corruption: Bribery" and ``Corruption: Other forms of corruption" were not included into the analysis as ``Corruption" variable has been computed by aggregating these two ones. Thus for this crime type, the rest categories in our dataset are treated as independent variables. 

Data of the ``Intentional Homicide" dataset, it was filtered for ``Total" dimensions, categories, sex and age groups only. The crime type was characterized with two indicators: ``Victims of intentional homicide" and ``Person convicted for intentional homicide". Using these indicators enabled the consideration of trends in the patterns of victimization for all categories, with respect to intentional homicide.

In the ``Violent and Sexual Crime" dataset, the indicator ``Violent offences" was selected, with the dimensions filtered by type of offence, while sex and age groups were aggregated under the total category. Within the ``Category" column, the values ``Sexual violence" and ``Other acts of sexual violence" were excluded. The first was excluded because its value depended on other features in the dataset, the second was also excluded because there were a high proportion of missing values that could potentially create biases \cite{jakobsen2017and}. With these exclusions, the remaining categories on the ``Category" column now became the variables that captured violent and sexual crimes.

During data preprocessing, separate observations representing the United Kingdom were aggregated into a single entry. The raw dataset had individual entries for ``United Kingdom (England and Wales)," ``United Kingdom (Northern Ireland)", and ``United Kingdom (Scotland)". These entries were combined into a single observation labeled as ``United Kingdom" by calculating the arithmetic mean of their respective values since all data were expressed in rates per 100,000 persons in the population. This guaranteed that the United Kingdom was considered as one unit in the analysis. 

The dataset excluded observations for Moldova, Vatican, Liechtenstein, and North Macedonia, since the number of missing values was large which can lead to potential bias in parameter estimation and diminish the generalizability of the findings \citep{schafer1997analysis,rubin2004multiple}. The reason to do so was to preserve the integrity and interpretability of the findings, as an overabundance of missing data may create biases or inaccuracies within the analytical framework.
 
The final data set consisted of variables taken from the three different data sources of crime; all measurements were in rates per 100,000 people. In order to prepare the data for CoDA, the variables were recognized as proportions to each country. 

\subsection{Imputation}
Compositional data, which are often found in various fields such as official statistics and environmental sciences, require special analysis methods due to their special characteristics. An observation is said to be a D-part composition when all its components comprise strictly positive real numbers and the information is contained in the ratios between these components \cite{aitchison1982statistical}. Standard statistical techniques, such as correlation analysis and principal component analysis, cannot be directly applied to compositional data because doing so often leads to spurious results \citep{pearson1897mathematical,aitchison1982statistical,filzmoser2009correlation}. This restriction holds in the same way for imputation methods \cite{bren2008news,martin2003dealing}.

In the dataset, there were \textit{32} missing values across the $33x14$  dimensions. In order to tackle the issue of absent values in compositional datasets, historical data was examined to aid in the imputation process. A linear regression model was applied between years and corresponding values, which allowed for the successful imputation of \textit{14} missing values using this approach. For those values we are unable to fill, iterative modeling techniques were utilized as outlined by Hron et al.\cite{hron2010imputation}. To begin with, the K-Nearest Neighbors (KNN) algorithm was used for the imputation of missing values. Subsequently, least trimmed squares (LTS) regression was implemented, recognized for its robustness and computational effectiveness \cite{rousseeuw2006computing}. This method was run 100 times, and the average of the results was taken for the sake of randomness. This methodology guaranteed that the imputation procedure was was both reliable and consistent with the compositional characteristics of the data. 

\subsection{Compositional Data and Aitchison Geometry}
\label{subsec:statanalys}
The analysis of compositional data requires the use of special methodologies because compositional data do not follow the
usual Euclidean geometry. Due to the peculiar properties of the data, which are constrained to the simplex space, traditional statistical methods cannot be directly applied in their original form. Therefore, it is necessary to adopt an appropriate geometrical framework for their analysis. Though Aitchison, in his landmark work in 1982, did not treat the geometrical aspect of the compositional data explicitly, later advancements \citep{pawlowsky2001geometric,egozcue2003isometric} introduced the concept of Aitchison geometry. In this approach, an attempt has been made to establish a vector space structure for the simplex to analyze compositional data effectively. Classic techniques in the field are based on traditional transformations like additive log-ratio (alr) and centered log-ratio (clr), as proposed by Aitchison \cite{aitchison1982statistical}. However, developments during the past two decades have pointed out the importance of employing isometric log-ratio (ilr) coordinates, which express the simplex as an Euclidean space, thus providing a more solid framework for the analysis of compositional data \cite{egozcue2003isometric}. 

\subsection{Cluster Analysis}
\label{subsec:cluster_analysis}
Cluster analysis, a technique that divides data into unique subgroups in order to make the similarities of the members within each subgroup stronger while making the differences between the subgroups even greater \cite{rokach2005clustering} is applied to two main parts: observations and variables. The optimal number of clusters was determined using the elbow method \cite{thorndike1953belongs}, and silhouette coefficient \cite{rousseeuw1987silhouettes} which provides a balance between the within-cluster variance and the overall clustering quality. 
First, the data are transformed into isometric log-ratio (ilr) coordinates without scaling, since scaling should be strictly avoided for coordinates \cite{filzmoser2018applied}. Based on these coordinates, distance matrices were calculated and hierarchical clustering, were developed by Macnaughton-Smith et al.\cite{macnaughton1964dissimilarity} was performed in order to explore the hierarchical structure of the data. Several clustering algorithms were used, including k-means, to define groups of similar environmental conditions based on the input covariates and the chosen number of clusters in environmental space \cite{hartigan1979k}, and Gaussian Mixture Models (GMM), a parametric probability density function represented as a weighted sum of Gaussian densities \cite{reynolds2009gaussian}, to ensure a thorough evaluation of clustering solutions. 

\subsection{t-distributed Stochastic Neighbor Embedding (t-SNE)}
\label{subsec:tsne}
t-distributed Stochastic Neighbor Embedding (t-SNE) is a dimensionality reduction technique that visualizes high-dimensional data by converting similarities between data points into joint probabilities and minimizing the Kullback-Leibler divergence between these probabilities in the high-dimensional and low-dimensional spaces such as 2D or 3D  \cite{van2008visualizing}.
\begin{equation}
    \label{eq:tsne_qij}
q_{ij} = \frac{ \left( 1 + \| y_i - y_j \|^2 \right )^{-1} }
{ \sum_{k \ne l} \left( 1 + \| y_k - y_l \|^2 \right )^{-1} }
\end{equation}
where $y_i$ and $y_j$ are points in the low-dimensional map in Eq.~(\ref{eq:tsne_qij}). The heavy tails enable modeling of moderate distances in high-dimensional space to be represented by larger distances in the low-dimensional map, reducing misleading attractive forces between dissimilar points. The gradient of the Kullback-Leibler divergence $C$ with respect to the map points is given by Eq.~(\ref{eq:tsne_c}):
\begin{equation}
    \label{eq:tsne_c}
\frac{\partial C}{\partial y_i} = 4 \sum_j \left( P_{ij} - q_{ij} \right ) \left( y_i - y_j \right ) \left( 1 + \| y_i - y_j \|^2 \right )^{-1}
\end{equation}
This formulation allows t-SNE to capture both the local and global data structure, and it is suitable for the visualization of high-dimensional complex data sets in a two-dimensional or three-dimensional form. This method was applied to preserve both local and global structures, reveal patterns at different scales, and visualize and evaluate the clusters. This is done by using ilr coordinates coupled with Aitchison distance, as well as tuning hyperparameters like perplexity and learning rate to obtain the best performance. The t-SNE result was compared with the clustering outcomes in order to evaluate alignment and validate the robustness of the identified clusters. 


\subsection{Q-mode Clustering}
\label{subsec:q-mode}
The primary focus was on grouping the observations (R-mode clustering) in previous sections~ (\nameref{subsec:cluster_analysis} \& \nameref{subsec:tsne}) but, the objective in Q-mode clustering is to group the variables or compositional components \cite{filzmoser2018applied}. Before delving into the details of the method, it would be helpful to explain basic concepts of distance and variation in compositional data. However, there is no direct counterpart  for variation estimation in compositional data, in contrast to the establishment of a center for a composition in origin. Instead, one deals with source information in a composition, i.e., with pairwise logratios. This creates the variation matrix \cite{aitchison1982statistical}, calculated from variances of pairwise logratios. Specifically, for a compositional data matrix $X = (x_{ij})$, the variation matrix is defined as
\begin{equation}
\label{eq:variation_matrix}
\mathbf{T}=\left(\begin{array}{cccc}
t_{11} & t_{12} & \ldots & t_{1 D} \\
t_{21} & t_{22} & \ldots & t_{2 D} \\
\vdots & \vdots & \ddots & \vdots \\
t_{D 1} & t_{D 2} & \ldots & t_{D D}
\end{array}\right)
\end{equation}
where $t_{j k}, j, k=1, \ldots, D$, are sample variances of pairwise logratios between $x_{j}$ and $x_{k}$, i.e.
\begin{flalign*}
&t_{j k}=\frac{1}{n-1} \sum_{i=1}^{n}\left(z_{j k}^{i}-\bar{z}_{j k}\right)^{2} \hspace{1cm} with \\ 
&\left\{z_{j k}^{i}=\ln \frac{x_{i j}}{x_{i k}}, i=1, \ldots, n\right\} \hspace{1cm} and \hspace{1cm} \bar{z}_{j k}=\frac{1}{n} \sum_{i=1}^{n} z_{j k}^{i}
\end{flalign*}

The main elements for cluster analysis are the distances or dissimilarities. In this context, the variation matrix  serves as a measure of dissimilarity among variables, analogous to the distance metrics used previously for the observations. The fact that it is symmetric and with zeros on its diagonal allows direct use of its elements as a dissimilarity metric for clustering analysis \citep{van2013analyzing,mckinley2016single}. Nevertheless, it has to be noted that the variation matrix does not show the formal properties of a distance matrix, as was stated by Fa{\v{c}}evicov{\'a et al. \cite{favcevicova2016element}. There are robust and classical ways to calculate a variation matrix. One of the most well-known multivariate location and covariance estimators is the Minimum Covariance Determinant (MCD) estimator \cite{rousseeuw1999fast}. It identifies $h$ observations $\left\{\mathbf{x}_{i_{1}}, \ldots, \mathbf{x}_{i_{h}}\right\}$ with minimum determinant of the sample covariance matrix over all sets of $h$ observations. The location estimator of MCD, $\mathbf{t}_{\mathrm{MCD}}$, shown as Eq.~\ref{tmcd}, is calculated as the arithmetic mean of $h$ observations, and the covariance estimator of MCD, $\mathbf{C}_{\mathrm{MCD}}$, by their sample covariance, multiplied by a constant $c_{\mathrm{MCD}}$, shown as Eq.~\ref{cmcd}, for consistency under normality,
\begin{flalign}
\mathbf{t}_{\mathrm{MCD}} & =\frac{1}{h} \sum_{j=1}^{h} \mathbf{x}_{i_{j}} \label{tmcd}\\
\mathbf{C}_{\mathrm{MCD}} & =c_{\mathrm{MCD}} \cdot \frac{1}{h-1} \sum_{j=1}^{h}\left(\mathbf{x}_{i_{j}}-\mathbf{t}_{\mathrm{MCD}}\right)\left(\mathbf{x}_{i_{j}}-\mathbf{t}_{\mathrm{MCD}}\right)^{\prime} \label{cmcd}
\end{flalign}

$h$ must represent a majority of the data and can be taken as an integer in the interval $[(n+p+1) / 2, n]$. In practice, a common practice is $h \approx 0.75 \cdot n$.
A variation matrix was robustly estimated using the MCD estimator and provided the necessary input for the hierarchical clustering algorithm and then used as input for the hierarchical clustering algorithm with the Ward's method \cite{ward1963hierarchical} in order to identify meaningful groups of variables. 

\subsection{Principal Component Analysis (PCA)}
\label{subsec:pca}
Principal component analysis (PCA) \cite{pearson1901liii}, being the most frequent multivariate method used in simplifying the dataset through dimensionality reduction, was performed on the newly clusters created from the previous Section~(\nameref{subsec:q-mode} in order to enable further analysis of its structure and dimensional properties. It generates orthogonal linear latent variables called principal components (PCs), which are characterized by loading matrices, showing the coefficients of the linear combinations based on the original variables, and scores, the values of the principal components \citep{johnson2007applied,varmuza2016introduction}. Although there are various methods to estimate principal components, singular value decomposition (SVD) and the covariance matrix decomposition were preferred in this paper. For compositional data, a standard way of formulating PCA and its corresponding biplot display were based on clr coefficients, as suggested by Aitchison \cite{aitchison2002biplots}.

In order to introduce SVD method, let a compositional data set be represented in terms of $n \times D$ data matrix $\mathbf{X}$. Let $\mathbf{Z}$ denote mean-centered coordinate data matrix. Mean-centering is crucial here in order to obtain directions of maximum variance. SVD decomposes $n \times (D-1)$ matrix $\mathbf{Z}$ into three parts:
\begin{equation}
\label{eq:svd1}
\mathbf{Z}=\mathbf{U D W}^{\prime} 
\end{equation}
Here, $\mathbf{U}$ is an $n \times p$ orthogonal matrix with the left singular vectors, $\mathbf{D}$ is a $p \times p$ diagonal with positive singular values $d_1, \ldots, d_p$ and $\mathbf{W}$ is a $(D-1) \times p$ orthogonal with the right singular vectors. $p=\min (n, D-1)$ is a parameter representing largest number of principal components computable, and it is determined from smaller dimension of data matrix $\mathbf{Z}$. Re-writing Eq.~(\ref{eq:svd1}):
\begin{equation}
\mathbf{Z}=(\mathbf{U D}) \mathbf{W}^{\prime}=\mathbf{Z}^{*} \mathbf{W}^{\prime}
\end{equation}
illustrates the PCA transformation. $\mathbf{Z}^{*}=\left(z_{i j}^{*}\right)$ contains the coordinates of samples in new coordinate system, denoted as scores. The columns of $\mathbf{U}$ produce same scores in a normalized form with unit variances, and variances of columns in $\mathbf{Z}^{*}$ denote variances of respective principal components. The columns of the matrix $\mathbf{W}$ known as the right singular vectors, are denoted to as loadings, and the matrix $\mathbf{W}$ itself is termed the loading matrix.

Second approach to estimate PCA is decomposition of covariance matrix. Due to affine equivariance \cite{filzmoser2018applied} of MCD estimator, i.e., selection of ilr coordinates will not impact resulting principal components. Algorithm for calculation of principal components with use of MCD estimator (for $n \geq D-1$) is discussed in \cite{filzmoser2009principal}. Loading matrix is calculated through an eigenvalue decomposition of the MCD covariance matrix:
\begin{equation}
\label{eq:pca_mcd1}
\mathbf{C}_{\mathrm{MCD}}=\mathbf{W}_{\mathrm{MCD}} \mathbf{D}_{\mathrm{MCD}} \mathbf{W}_{\mathrm{MCD}}^{\prime}
\end{equation}
$\mathbf{W}_{\mathrm{MCD}}$ is an eigenvector matrix, and it can serve as a loading matrix, and $\mathbf{D}_{\mathrm{MCD}}$ is an eigenvalue (robust variances of principal components) diagonal matrix. To calculate principal component scores, first, coordinates matrix is robust mean-centered column-wise using elements of $\mathbf{t}_{\mathrm{MCD}}$. Robustly centered coordinates are denoted in $\mathbf{Z}_{\mathrm{MCD}}$. After Eq.~(\ref{eq:pca_mcd1}), following computation of PCA scores are then computed as:
\begin{equation}
\label{eq:pca_mcd2}
\mathbf{Z}_{\mathrm{MCD}}^{*}=\mathbf{Z}_{\mathrm{MCD}} \mathbf{W}_{\mathrm{MCD}}
\end{equation}
Such principal components resist outliers \cite{croux2000principal}. Robustness in such principal components ensures one obtains first principal components representing information contained in most of the data's joint distribution. On the contrary, non-robust PCA can become sensitive to outliers, and first principal components can point in directions according to such outliers and generate large (non-robust) variances. That is not preferable, as the goal of PCA is not to detect outliers but to reduce dimensionality by summarizing the majority of the data's information. PCA was performed for both classical and robust methods and the one with the higher explained variance ratio was preferred. This approach did indeed allow for effective interpretation of the underlying structure of the data in view of its compositional nature and identification of the dominant patterns within each cluster, showing shared variability among the grouped variables. The current analysis combines Q-mode clustering and PCA, therefore giving insight into the relationships between variables and their contribution to the overall structure of the dataset.

\section{Results}
\label{sec:results}
The results of this study provide a solid foundation for a complete understanding of the clustering behavior of European countries based on their crime rates in 2022. This study, using k-means clustering, hierarchical clustering, and Gaussian Mixture Models (GMM), found distinct groupings of countries, hence serving to give insight into regional similarities. In addition, dimensionality reduction techniques, such as t-SNE and PCA, are applied in order to enhance the clustering results interpretability.

\subsection{Part 1: Examination of Observations}
\subsubsection{Cluster Analysis}
The analysis is done to comparatively assess different clustering algorithms to determine the most effective method for grouping countries. Of the three, k-means clustering with a parameter of 3 clusters proved to be the best approach to this analysis. The results of the clustering analysis, as presented in Table~\ref{tab:clustering_result}, assign each country to one of three identified clusters. 

\begin{longtable}{|c|c|c|c|} 
\caption{\bf Kmeans Clustering Results\label{tab:clustering_result}} \\
\hline
\textbf{\#} & \textbf{Iso3\_code} & \textbf{Cluster} & \textbf{Country} \\
\hline
\endfirsthead
\multicolumn{4}{c}%
{\tablename\ \thetable\ -- \textit{Continued from previous page}} \\
\hline
\textbf{\#} & \textbf{Iso3\_code} & \textbf{Cluster} & \textbf{Country} \\
\hline
\endhead
\hline \multicolumn{4}{r}{\textit{Continued on next page}} \\
\endfoot
\hline
\endlastfoot
1 & ALB & 2 & Albania \\
2 & AUT & 3 & Austria \\
3 & BIH & 2 & Bosnia and Herzegovina \\
4 & BGR & 2 & Bulgaria \\
5 & HRV & 3 & Croatia \\
6 & CZE & 3 & Czechia \\
7 & DNK & 3 & Denmark \\
8 & EST & 3 & Estonia \\
9 & FIN & 3 & Finland \\
10 & FRA & 1 & France \\
11 & DEU & 1 & Germany \\
12 & GRC & 2 & Greece \\
13 & HUN & 3 & Hungary \\
14 & ISL & 3 & Iceland \\
15 & IRL & 1 & Ireland \\
16 & ITA & 1 & Italy \\
17 & LVA & 2 & Latvia \\
18 & LTU & 3 & Lithuania \\
19 & LUX & 1 & Luxembourg \\
20 & MLT & 3 & Malta \\
21 & MNE & 2 & Montenegro \\
22 & NLD & 2 & Netherlands \\
23 & NOR & 1 & Norway \\
24 & POL & 2 & Poland \\
25 & PRT & 1 & Portugal \\
26 & ROU & 2 & Romania \\
27 & SRB & 2 & Serbia \\
28 & SVK & 2 & Slovakia \\
29 & SVN & 3 & Slovenia \\
30 & ESP & 1 & Spain \\
31 & SWE & 3 & Sweden \\
32 & CHE & 3 & Switzerland \\
33 & GBR & 1 & United Kingdom \\
\end{longtable}

In order to better contextualize these findings, Fig.~\ref{fig:countries_map} provides a spatial visualization of how the three clusters are distributed geographically. This map provides a geographical perspective on the clustering patterns found, and it clearly shows that the clusters align with regional similarities, which can result from related economic, cultural, or geographical characteristics.

\begin{figure}[!t]
    \includegraphics[width=13cm]{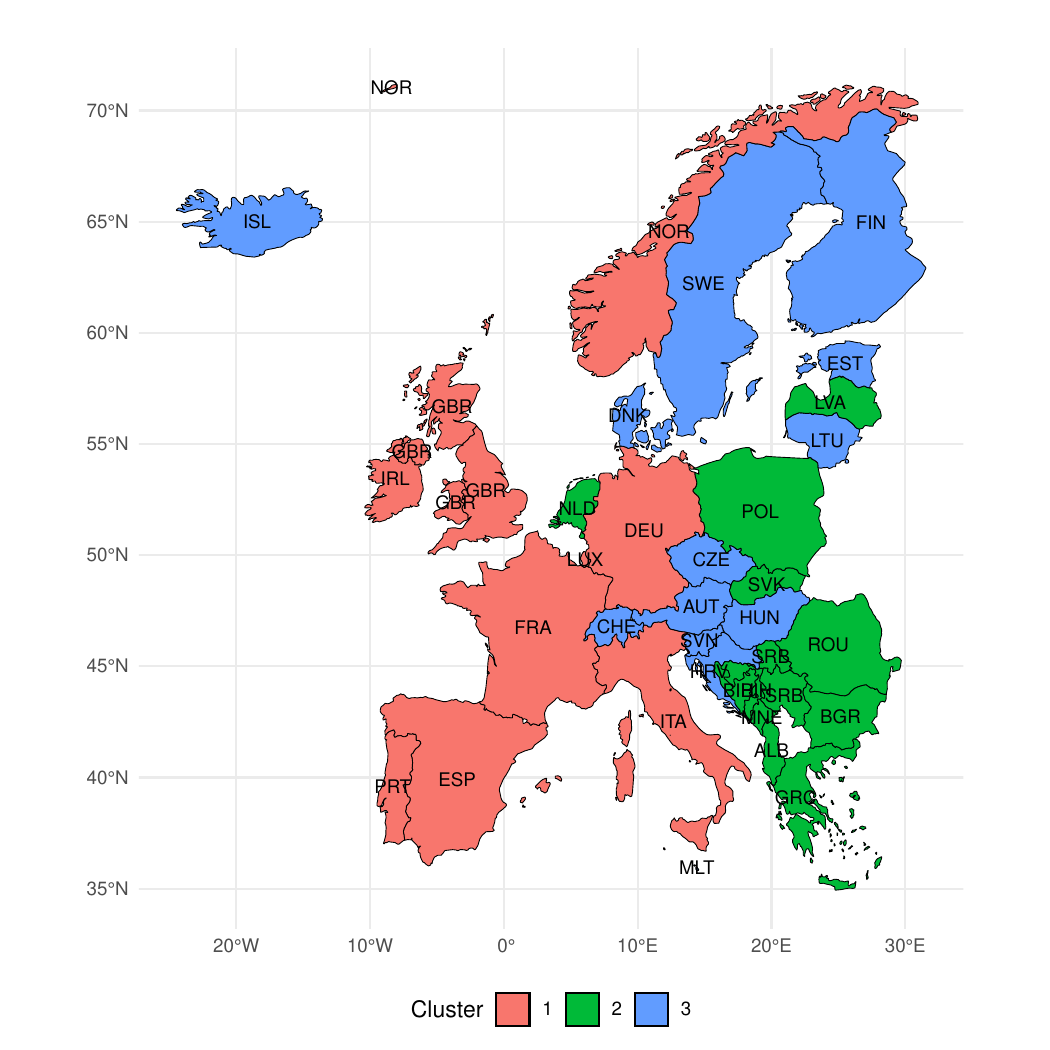}
    \caption{\bf Map of European Countries with Clustering Results}
    \label{fig:countries_map}
\end{figure}

Cluster 1 includes countries from West and South Europe, namely France, Germany, Italy, Portugal, and Spain, along with Ireland, Luxembourg, and Norway. These countries have high levels of economic development and very strong industrial bases. Notably, Norway, a Nordic country, is clustered here rather than with other Nordic countries in Cluster 3. Cluster 2 contains countries from Eastern and Southeastern Europe, such as Albania, Bulgaria, Romania, and Poland; this is reflective of shared historical and economic characteristics, including post-socialist transitions. Latvia—normally clustered with its Baltic neighbors in Cluster 3—is here, perhaps in view of some particular differences in the sets of economic and demographic indicators. Similarly, the Netherlands, which is often considered part of Western Europe, shows up in Cluster 2, interestingly. Cluster 3 encompasses countries from Northern and Central Europe, including Denmark, Sweden, Finland, Austria, and Switzerland, in addition to Estonia and Lithuania. These nations exhibit significant economic advancement and robust social welfare frameworks. Although the clustering largely corresponds with geographic areas, anomalies such as Latvia, Norway, and the Netherlands underscore the intricacies of the dataset and the specific variances inherent to the variables examined.

\subsubsection{Part 1: t-SNE Dimension Reduction}
To further investigate the relationships between countries, t-SNE had been used to reduce the dimensions of data for a visual representation of the clustering result. Fig.~\ref{fig:tsne_result} illustrates the result of two-dimensional t-SNE of the data. Points are colored according to their cluster assignment given by k-means.  Such a projection obtained by reduction of dimensions provides an intuitive insight into the relationships of countries, preserving at the same time an important property of the local structure in this dataset. Again, it is reassuring to see that the clusters are distant in t-SNE space, which further validates the results provided by k-means. Some minor intersections or scatter-for example, the positioning of Latvia and Norway-reflect the intrinsic complexity of the dataset and cannot destroy the overall coherency of the clustering. The visualization is a good means for interpretation of clustering results beyond numerical quantitative metrics and provides a clear understandable view of the pattern hidden in the data.

\begin{figure}[!t]
    \includegraphics[width=12cm,keepaspectratio]{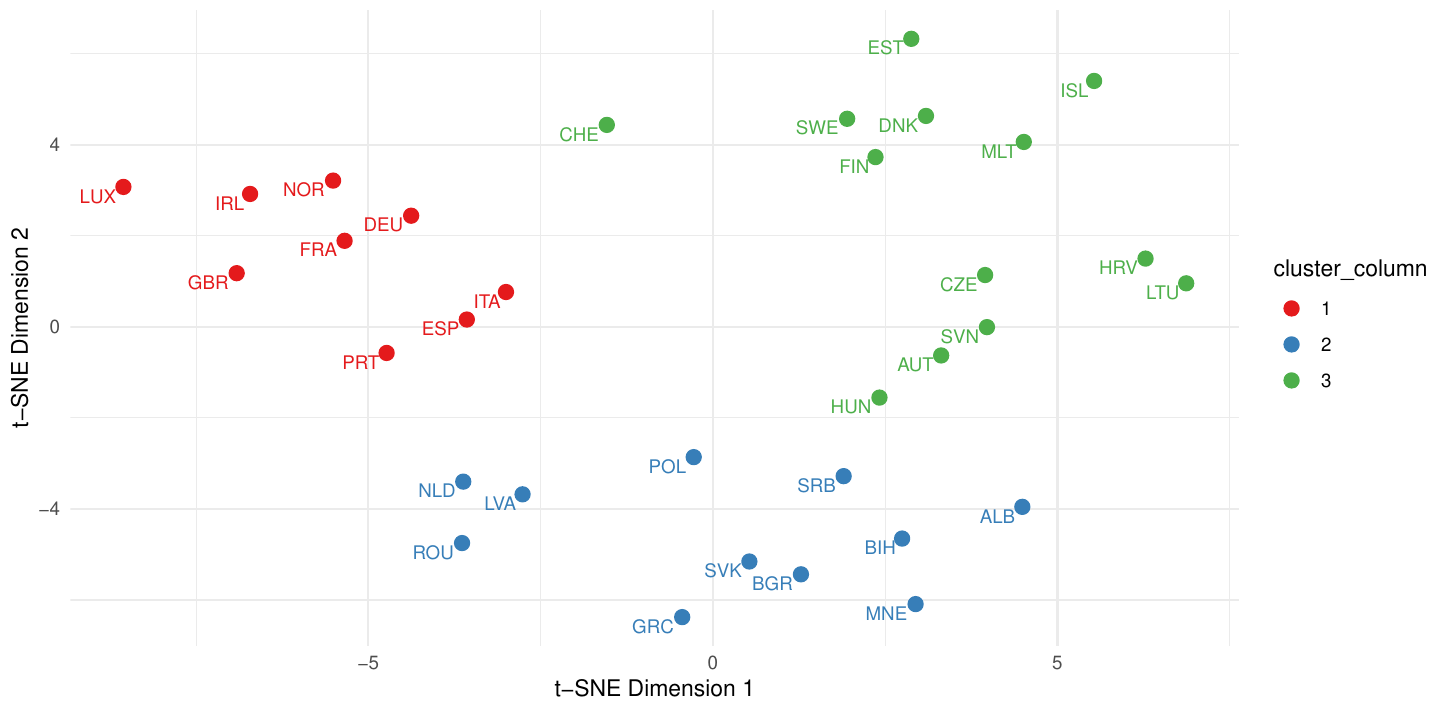}
    \caption{\bf t-SNE 2D Visualization Colored by KMeans}
    \label{fig:tsne_result}
\end{figure}

\subsection{Part 2: Clustering of Variables}
Fig.~\ref{fig:cluster_variables} shows the dendrogram of Ward method clustering applied to the robust variation matrix of the variables. The analysis identified three main clusters, each representing groupings of variables in variation patterns. These clusters will then be of service in the visualization of the relationships that will settle and group the variables into further analysis and interpretation. 

\begin{figure}[!t]
    \includegraphics[width=14cm,keepaspectratio]{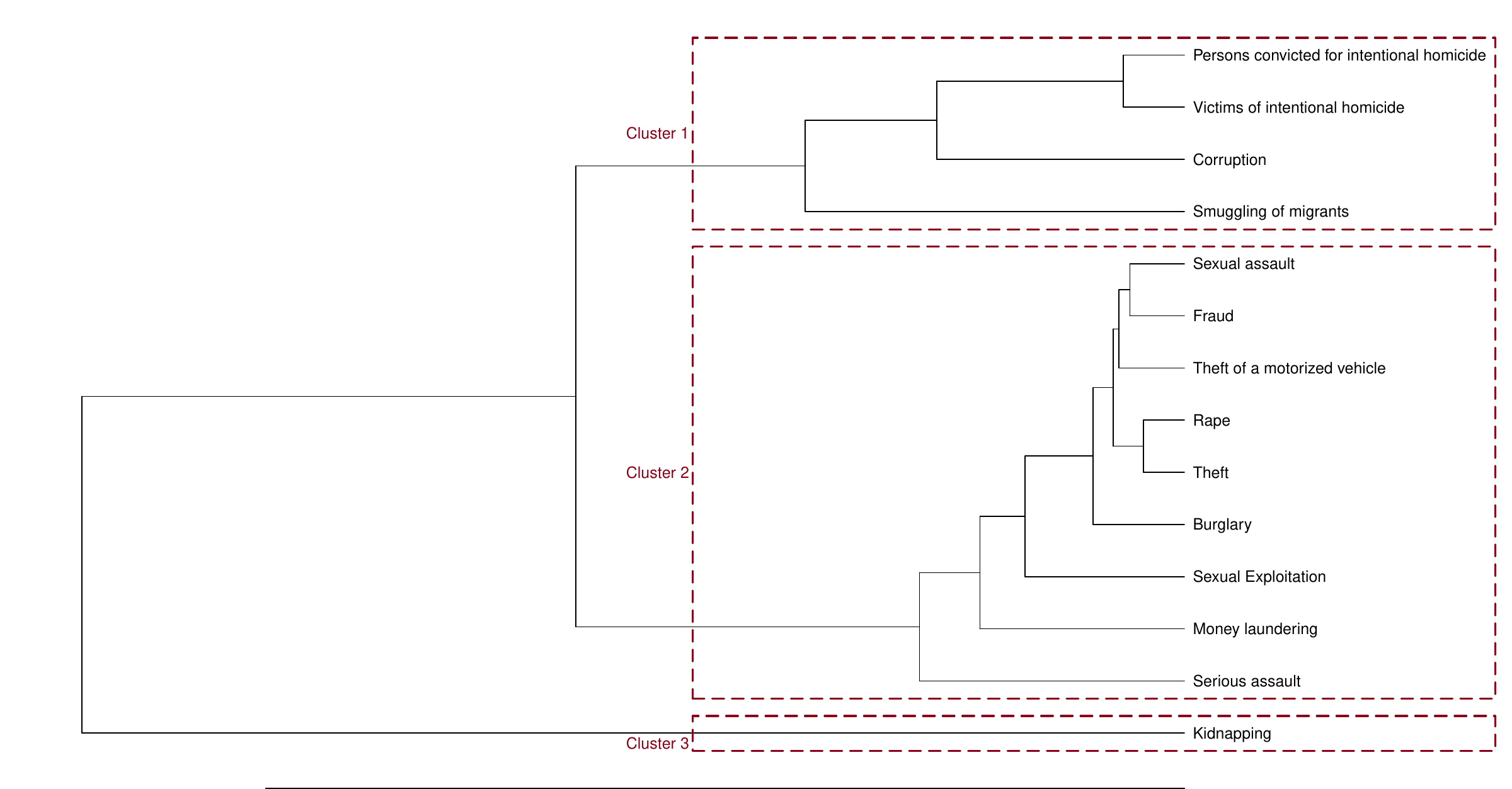}
    \caption{\bf Dendogram: Hierarchical Clustering Variables}
    \label{fig:cluster_variables}
\end{figure}

The PCA biplots, Fig.~\ref{fig:pca_cluster1_variables}, emphasize the relationships between four variables of Cluster 1. \textit{``Corruption", ``Smuggling of migrants", ``Victims of intentional homicide", and ``Persons convicted for intentional homicide"} — and their influence on PC1 and PC2. The robust method has a higher explained variance ratio (0.90) than the classical approach (0.87) and is therefore preferable. PC1 contrasts homicide-related variables — victims and convictions — with corruption and smuggling. Countries such as Austria (2) and Iceland (14) have a high score on PC1, showing higher corruption and smuggling, whereas countries such as Luxembourg (19) and Ireland (15) show stronger homicide-related factors. PC2 separates corruption from smuggling, with Estonia (8) exhibiting high corruption and Luxembourg (19) and Ireland (15) being more related to smuggling. 

\begin{figure}[!t]
    \includegraphics[width=14cm,keepaspectratio]{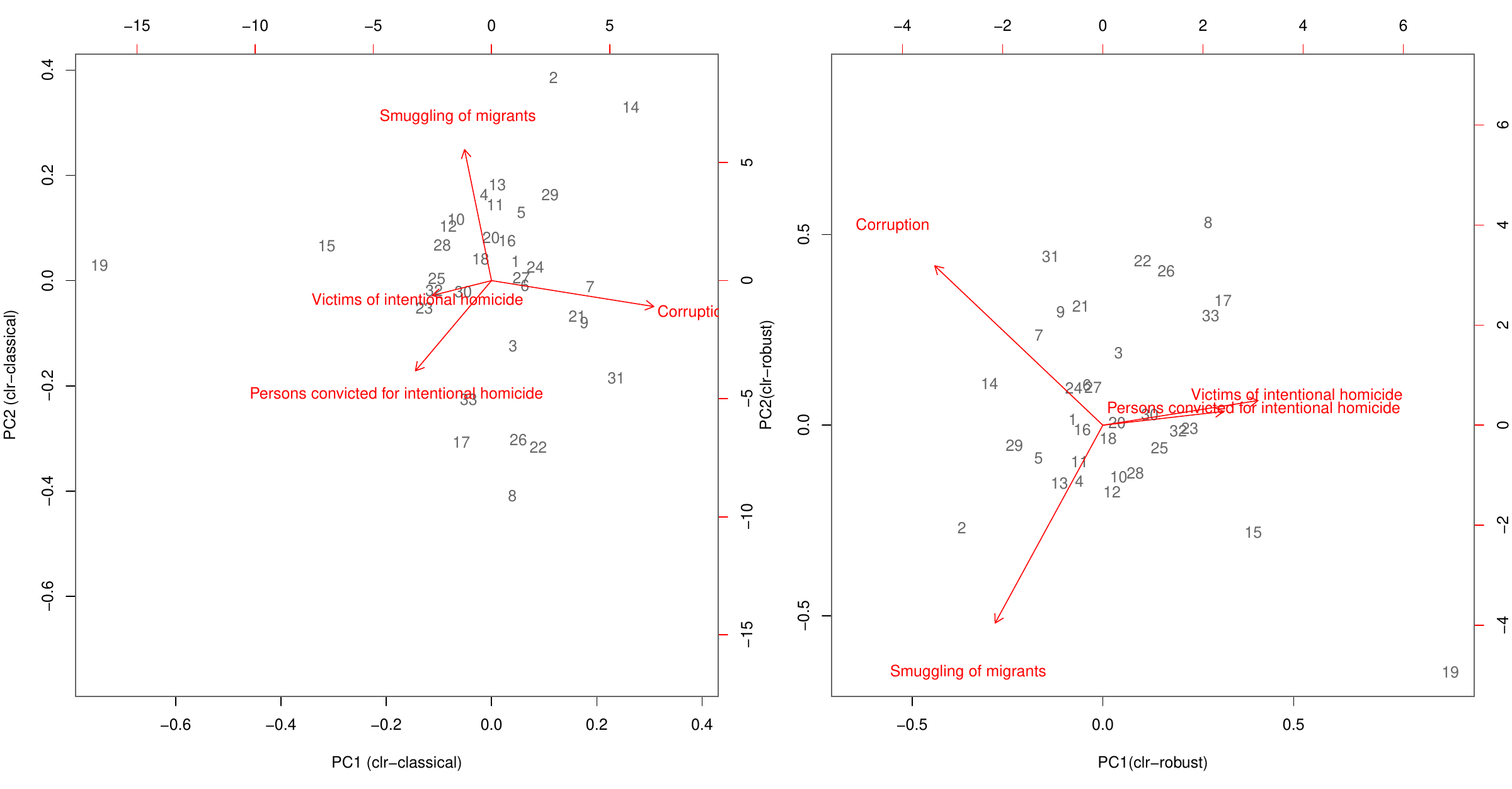}
    \caption{\bf Classical and Robust Biplots of Cluster 1 Variables}
    \label{fig:pca_cluster1_variables}
\end{figure}

The robust PCA methodology is clearly preferable to the classical method, with a ratio of 0.77 as against 0.63. Robust principal component analysis on the second cluster of variables, as shown in Fig.~\ref{fig:pca_cluster2_variables}, covers crimes like burglary (1), theft (2), fraud (4), and sexual crimes (8), and reveals some interesting patterns of criminal behavior across different countries. PC1 highlights a contrast money laundering (5) with crimes like serious assault (6), theft of a motorized vehicle (3), and sexual assault (8) in countries like Albania (1) and Denmark (7), which show a higher money laundering activity, while Luxembourg (19), Spain (30) and Italy (16) show a stronger focus on physical and property crimes, including theft, burglary, and serious assault. PC2 emphasizes serious assault (6) as a dominant factor, which puts it in opposition to fraud (4) and sexual exploitation (9). Countries such as Bosnia and Herzegovina (3) and Serbia (27) manifest a closer association with serious assault and violent crimes, whereas Romania (26) and Estonia (8) incline more towards financial crimes, represented by fraud, as well as exploitative crimes such as sexual exploitation.

\begin{figure}[!t]
    \includegraphics[width=14cm]{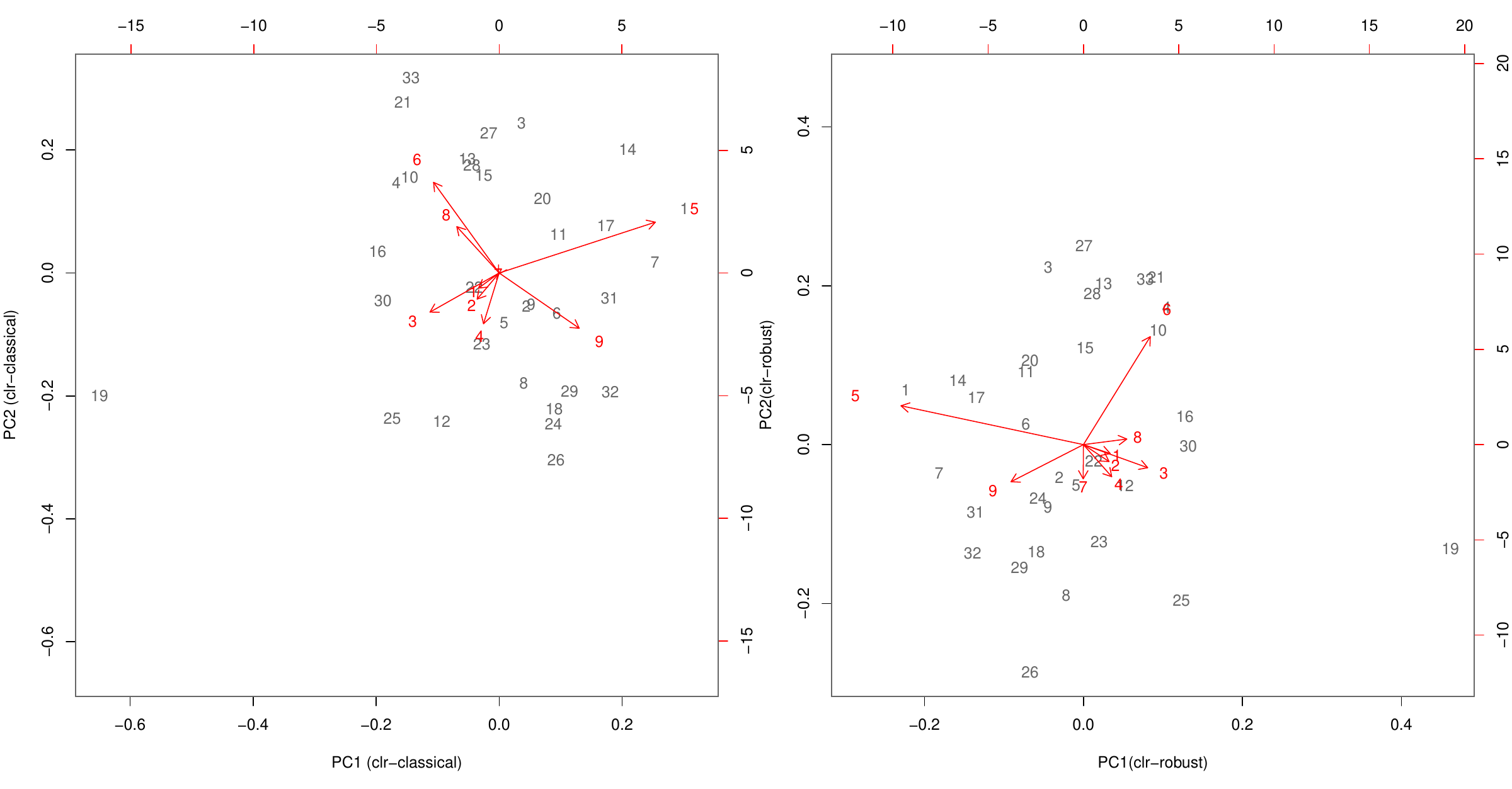}
    \caption{\bf Classical and Robust Biplots of Cluster 2 Variables}
    \label{fig:pca_cluster2_variables}
\end{figure}

\section{Conclusion and Discussion}
\label{sec:conclusion}

This study has demonstrated the efficacy of compositional data analysis in elucidating complex patterns within European crime data for the year 2022 to reveal insightful patterns and regional similarities. In this way, the implementation of k-means clustering returned three clear clusters separating the countries, basically by their geographical position and socio-economic characteristics. Moreover, t-SNE and geographic mapping underlined the fact that the three clusters were well separated. Outliers such as the clustering of Norway, Latvia, and the Netherlands were intriguing, yet again indicative of the very complex interaction of variables affecting the crime rate. The variable clusters unveiled through Robust PCA explained the relation between the different types of crimes. In this way, the analysis was able to show the connection between certain types of crimes-for example, homicide, smuggling, and financial crimes-and how they vary between countries. Specifically, homicides are most concentrated in regions with high levels of organized crime or socio-economic instability, at the same time supporting a range of illegal activity, such as smuggling and financial offences. All these processes depend on a complex interrelationship between geographical, socio-economic, and governance factors, with significant country and region-specific variances.

Nevertheless, the limitations of the present study need to be taken into consideration. First, it includes data only for 2022; therefore, neither long-term tendencies nor impacts of single events are reflected in it. The data that this study is based on is of reported crimes, which might carry certain biases because of the divergent conventions and practices of reporting and recording in different legal systems. Longitudinal data with more socio-economic variables, and possible biases in reporting, would add to this branch of research. Besides lack of recording for some countries or incomplete recording for some heads of crime, such analysis tends to get undermined. The fact that data reflect shared concepts and clear definitions \cite{harrendorf2010international}, and standardization of methods of data collection would enhance international comparability and further international efforts.

The power of compositional data analysis in unscrambling the hidden structures within crime data has been shown in this paper and will go on to prove useful insight for policymakers and researchers in times to come. Further research could thus be directed toward establishing causality, making use of more variables, and observing the changes in a trend over time.

\section{Competing interests}
No competing interest is declared.

\section{Author contributions statement}
All authors contributed equally to this paper.

\section{Data Availability Statement}
Publicly available data, along with the processed and imputed form of the data and the corresponding codes are available at the public repository \url{https://github.com/onurbdogan/crime_in_proportions_europe_2022}.

\section{Acknowledgments}
The authors thank the anonymous reviewers for their valuable suggestions.



\end{document}